\documentclass[letterpaper, 10 pt, conference]{ieeeconf}  

\IEEEoverridecommandlockouts                              
\usepackage{graphics} 
\usepackage{epsfig} 
\usepackage{mathptmx} 
\usepackage{amsmath} 
\usepackage{amssymb}  
\usepackage{algorithm}
\usepackage{algorithmic}
\usepackage{cite}
\newtheorem{myDef}{Definition}
\newtheorem{myRek}{Remark}
\newtheorem{myTheo}{Theorem}
\newtheorem{myLem}{Lemma}
\newtheorem{myEg}{Example}

\title{\LARGE \bf
Extended Insertion Functions for Opacity Enforcement
}

\author{Xiaoyan Li, Christoforos N. Hadjicostis, and Zhiwu Li
\thanks{This work was supported in part by the National Natural
Science Foundation of China under Grant 61873342, the
Science and Technology Development Fund (FDCT), MSAR SAR under Grant
0012/2019/A1, and the China Scholarship Council.}
\thanks{X. Y. Li is with the School of Electro-Mechanical Engineering, Xidian University, Xi'an 710071, China
        {\tt\small lixiaoyan@stu.xidian.edu.cn}}%
\thanks{C. N. Hadjicostis is with the Department of Electrical and Computer Engineering, University of Cyprus, Nicosia, Cyprus, and also with the School of Electro-Mechanical Engineering, Xidian University, Xi'an 710071, China
        {\tt\small chadjic@ucy.ac.cy}}%
\thanks{Z. W. Li is with the School of Electro-Mechanical Engineering, Xidian University, Xi'an 710071, China, and also with the Institute of Systems Engineering, Macau University of Science and Technology, Taipa 999078, Macau, China
        {\tt\small zhwli@xidian.edu.cn}}%
}

\begin{document}

\maketitle
\thispagestyle{empty}
\pagestyle{empty}

\begin{abstract}
Opacity is a confidentiality property that holds when certain secret strings of a given system cannot be revealed to an outside observer under any system activity.
Opacity violations stimulate the study of opacity enforcement strategies.
Among other methodologies, opacity has been enforced using insertion mechanisms, i.e., output obfuscation mechanisms that are allowed to insert fictitious output symbols before actual system outputs, in order to preserve opacity.
This paper studies and analyzes more powerful extended insertion mechanisms, which can insert symbols before and after an actual system output, thus, providing opacity to a wider class of systems.
In order to address practical considerations, the paper also introduces event insertion constraints (i.e., the case when only specific symbols can be inserted before and/or after an actual system output).
For each case, we construct an appropriate verifier that can be used to obtain necessary and sufficient conditions for checking opacity enforceability.
\end{abstract}

\section{INTRODUCTION}

Security and privacy of cyber-physical systems are increasingly emphasized nowadays considering that network infrastructures, the underlying mechanisms for exchanging information within these systems, are threatened and can become compromised by malicious entities.
A hostile entity can selectively attack network infrastructure and cause immeasurable damage to a system in terms of privacy violations.
Opacity, initially proposed in the computer science community \cite{opacity-presented}, can be used to capture important security and privacy properties of an underlying system.
Opacity has also been considered in discrete event systems, such as Petri nets and automata, to verify and enforce security and privacy properties in dynamically evolving environments \cite{Introduction-DES,chris-book,tong-2017-v,saboori2011verification1,two-opacity-verify,chris-2014,chris-planning}.

In the context of discrete event systems, opacity is a property that holds when some predefined secrets (e.g., certain sequences of events or certain states) of the system can be kept secret from external observers/intruders \cite{Falcone}.
Opacity can be seen as an information-flow property from the system/defender to the intruder. More specifically, the system generates an observation (or a sequence of observations), and the intruder aims to determine whether the observation (or the sequence of observations) has necessarily been generated by secret behavior.
Typically, if the secrets cannot be definitely inferred by the intruder (under any underlying system behavior), then the system is considered to be opaque.

When opacity is violated, there are four major methods to enforce it: supervisory control, dynamic observability, insertion functions and edit functions\cite{dynamic-observability,tong-2018-e,SC-opacity1,SC-opacity2,SC-opacity3,IF-opacity1,IF-opacity2,IF-opacity3,
EIF-opacity,MIF-opacity,EF-1,PPIF-opacity}.
This paper extends the insertion function in \cite{IF-opacity1} by investigating extended insertion functions and their application to opacity enforcement.
We also consider a scenario in terms of how the extended insertion mechanism might be constrained. More specifically, we consider event insertion constraints (EICs), i.e., situations where only certain symbols can be inserted before and/or after the observed event.

The contribution of this paper is twofold. The first contribution concerns the introduction of the notions of EI-enforceability and EIC-enforceability.
A system is EI-enforceable if opacity can be enforced via the proposed extended insertion mechanism; and
a system with EICs is EIC-enforceable if opacity can be enforced via the proposed extended insertion mechanism by inserting only specific types of events that are allowed.
The second contribution concerns the construction of a verifier, and the development of necessary and sufficient conditions for checking the feasibility of these two methods of enforcing opacity.

The remainder of this paper is arranged as follows.
Preliminaries are recalled in Section~\ref{background} embracing basic notions, such as system model, opacity, and insertion functions.
Section~\ref{enforceability} defines extended insertion sequences as well as EI-enforceability and EIC-enforceability.
In Section~\ref{enforceability-verification}, the verifier and EIC-verifier constructions are reported and two necessary and sufficient conditions for EI-enforceability and EIC-enforceability are developed. Finally, Section~\ref{conclusions} concludes the paper.

\section{PRELIMINARIES AND BACKGROUND}\label{background}

\subsection{Basic Notions and System Model}

We use $E$ and $E^{*}$ to denote an alphabet and the set of all finite-length strings of elements of $E$, respectively.
Given a string $s\in E^{*}$, its length, denoted by $|s|$, represents the number of events in $s$ (the empty string $\epsilon$ is also a member of $E^{*}$ and its length is zero).
Given two strings $s, t\in E^{*}$, the concatenation of $s$ and $t$, denoted by $st$, indicates that the event sequence captured by $t$ immediately happens after the occurrence of the event sequence captured by $s$.
For a given string $s\in E^{*}$, $s^{*}$ represents the set of strings obtained by concatenating a finite number (but possibly arbitrarily large) of $s$ (including $\epsilon$ when the number of times we concatenate is zero).
String $s'\in E^{*}$ is called a prefix of $s$ if there exists $s''\in E^{*}$ such that $s=s's''$, and the set of all prefixes of $s$ is denoted by $\overline{s}$.
We use $|A|$ to denote the number of elements in a set $A$.

A deterministic finite automaton (DFA) is a four-tuple ${G_d=(X, E, f, x_0)}$, where $X$ is the set of states, $E$ is the set of events, $f: X \times E \rightarrow X$ is the possibly partially defined deterministic transition function, and $x_0 \in X$ is the initial state. The transition function $f$ can be extended from $X \times E$ to $X \times E^{*}$ in a recursive way, which is defined as $f(x, es)=f(f(x, e), s)$, where $e\in E$ and $s\in E^{*}$ (specifically, $f(x, es)$ is undefined if $f(x, e)$ is undefined and we have $f(x, \epsilon)=x$ for all $x\in X$).
The behavior of a given DFA is captured by its generated language, defined as $L(G_d)=\{s\in E^{*}|f(x_0, s)$ is defined$\}$.
For a state $x\in X$, we use $T(x)=\{e|f(x, e)$ is defined$\}$ to denote the set of transitions that can occur at $x$, and $F(x)=\{e|\exists x'\in X\{f(x', e)=x\}\}$ to denote the set of transitions that can reach $x$.

A nondeterministic finite automaton (NFA) is a four-tuple ${G_{nd}=(X, E, \delta, X_0)}$, where $X$ is the set of states, $E$ is the set of events, $\delta: X \times E \rightarrow 2^X$ is the nondeterministic transition function, and $X_0 \subseteq X$ is the set of possible initial states.
The transition function $\delta$ can be extended from $X \times E$ to $X \times E^{*}$ in a recursive way, which is defined as $\delta(x, es)=\bigcup_{x'\in\delta(x, e)}\delta(x', s)$, where $e\in E$ and $s\in E^{*}$ (specifically, $\delta(x, \epsilon)=\{x\}$ for all $x\in X$). The behavior of a given NFA is captured by its generated language, which is defined as $L(G_{nd})=\{s\in E^{*}|\exists x_0\in X_0\{  \delta(x_0, s)\neq \emptyset\}\}$.

In this paper, we consider all events of a DFA as observable. For an NFA, we assume that some events are observable and the remaining events are unobservable.
Thus, the set of events of an NFA can be partitioned into two parts: $E=E_{o}\cup E_{uo}$ ($E_{o}\cap E_{uo}=\emptyset$), where $E_{o}$ is the set of observable events and $E_{uo}$ is the set of unobservable events.
The natural projection of an event is defined as $P(e)=e$ if $e\in E_o$, otherwise $P(e)=\varepsilon$ if $e\in E_{uo}$.
Given a string $s\in E^{*}$, its observation is the output of the natural projection $P: E^{*}\rightarrow E_o^{*}$, which is defined recursively as $P(s)=P(s'e)=P(s')P(e)$ for $s=s'e$, where $s'\in E^{*}$ and $e\in E$.

An NFA $G_{nd}=(X, E_o\cup E_{uo}, \delta, X_0)$ can be converted into a DFA $G_{d}$ (with all events observable) by obtaining the observer of $G_{nd}$ that is denoted by $Obs(G_{nd})=(X_{obs}, E_o, f_{obs}, x_{0, obs})$ according to the method in \cite{Introduction-DES}.
More specifically, $X_{obs}\subseteq 2^X$ is the state space, $E_o$ is the set of observable events, $f_{obs}: X_{obs}\times E_o\rightarrow X_{obs}$ is the transition function, and $x_{0, obs}$ is the initial state.
Note that each observer state $x_{obs}\in X_{obs}$ is associated with a subset of states obtained from $X$, i.e., $x_{obs}\subseteq X$, and we call $x_{obs}\in X_{obs}$ the current state estimate of system $G_{nd}$, sometimes abbreviated by simply estimate.
Consider an NFA $G_{nd}=(X, E_o\cup E_{uo}, \delta, X_0)$. Let $Obs(G_{nd})=(X_{obs}, E_o, f_{obs}, x_{0, obs})$ be the observer. Given a string $s\in L(G_{nd})$, the set of all possible states following the observation $P(s)$ is $R(x_{0, obs}, P(s))=\{x\in X|\exists x'\in x_{0, obs}, \exists s'\in L(G_{nd}), s.t.\ \{P(s')=P(s)\wedge x\in \delta(x', s')\}\}$.
Based on the observations, all estimates can be obtained by tracking the sequence of observations generated by the system (the initial estimate is $x_{0, obs}=R(X_0, \varepsilon)$) \cite{chris-book}.

A DFA ${G_d=(X, E, f, x_0)}$ can also be viewed as a directed graph with $X$ being the set of nodes and $\{(x, f(x, e))|f(x, e)$ is defined\} being the set of directed edges.
The postset of $x$ is defined as $x^\bullet=\{x'\in X|\exists e\in E \{x'=f(x, e)\}\}$.
A sequence $x_1x_2...x_n$, composed of $n$ nodes, is said to be a path, denoted by $p: x_1\Rightarrow x_n$, if for all $i\in \{1, 2, ..., n-1\}$ $x_{i+1}\in x_i^\bullet$.
A directed graph $G$ is strongly connected if for each pair of nodes $x, x'$ in $X$ we have paths $p_1: x\Rightarrow x'$ and $p_2: x'\Rightarrow x$.
A subgraph $G'$ of $G$ is called a strongly connected component (SCC) if $G'$ is strongly connected and there is no additional node from $G$ that can be included in $G'$ without breaking the property of being strongly connected. (Note that a single node $x$ can also be an SCC if there is no other node that can form an SCC with $x$).
The set of nodes of an SCC is denoted by $X^{SCC}$, which can also be used to represent the corresponding SCC.

\subsection{Opacity and Insertion Functions for Opacity Enforcement}

\begin{myTheo}\label{current-state-opacity}(Current-state opacity \cite{chris-book})
Given an NFA $G_{nd}=(X, E_o\cup E_{uo}, \delta, X_0)$ with $X_S\subseteq X$ being the set of secret states and $Obs(G_{nd})=(X_{obs}, E_o, f_{obs}, x_{0, obs})$ being the corresponding observer (constructed with respect to the set of observable events $E_o$), $G_{nd}$ is current-state opaque if $$\forall x_{obs}\in X_{obs} \{\exists x\in (X\backslash X_S)\{x\in x_{obs}\}\}.$$
\end{myTheo}

An estimate $x_{obs}\in X_{obs}$ is called {\em secret} or  {\em unsafe} if for all $x\in x_{obs}$, $x\in X_S$. In the remaining contents, we consider an estimate as an observer state and a secret estimate as a secret (observer) state.
Given a DFA $G=(X, E, f, x_0)$ with $X_S\subseteq X$ being the set of secret states, the behavior of the system $L(G)$ can be partitioned into two disjoint sets: current state-safe behavior $L_{cs}(G)$ and current state-unsafe behavior $L_{cus}(G)$, which are defined respectively as
$L_{cs}(G)=\{t|t\in L(G) \wedge (f(x_0, t)\in (X\backslash X_S))\}$ and $L_{cus}(G)=L(G)\setminus L_{cs}(G)$, respectively.
If a system generates a current state-unsafe observation, the fact that the state of the system is secret is certain to be revealed.

If a system is nonopaque, the insertion functions proposed by Wu \emph{et al.} \cite{IF-opacity1} can be used to attempt to enforce opacity. To enforce opacity, the system defender needs to make the unsafe observation look like safe observation by using insertion functions. The insertion function is applied by inserting an observable event sequence before the actual observation. The decision as to what to insert needs to be made based on the actual observation (as seen last) as well as the previous observations and the previous insertion actions that have taken place.
To distinguish between events actually observed versus events virtually inserted, we define the set of inserted events $E_i=\{e_i|e\in E\}$ for a DFA.
From the perspective of an intruder, $e_i$ looks like $e$.
Given an observation $s=s[1]s[2]...s[n]\in L(G_{d})$ whose length is $n$, an insertion sequence is denoted by $s_I(s)=s_i[1]s_i[2]...s_i[n]$, where $s[m]$ represents an actual observable event, $s_i[m]\in E_{i}^{*}$ indicates the inserted observable event sequence associated with $s[m]$, and $m\in \{1, 2, ..., n\}$ is a positive integer. After implementing $s_I(s)$ on $s$, the modified observation $s_M(s)=s_i[1]s[1]s_i[2]s[2]...s_i[n]s[n]$ captures what is observed by the intruder.

We make the worst-case assumption that the intruder has perfect knowledge of a given system, which means that the intruder can distinguish the authenticity of the received observation according to the transition structure of the system.
The intruder will become suspicious of any observation that cannot be generated from the system.
To avoid raising suspicion to the intruder, it is essential for the defender to generate modified observation that matches valid observation, which can indeed be generated by the system.

\begin{figure}[b]
\centering
\includegraphics[height=4.4cm]{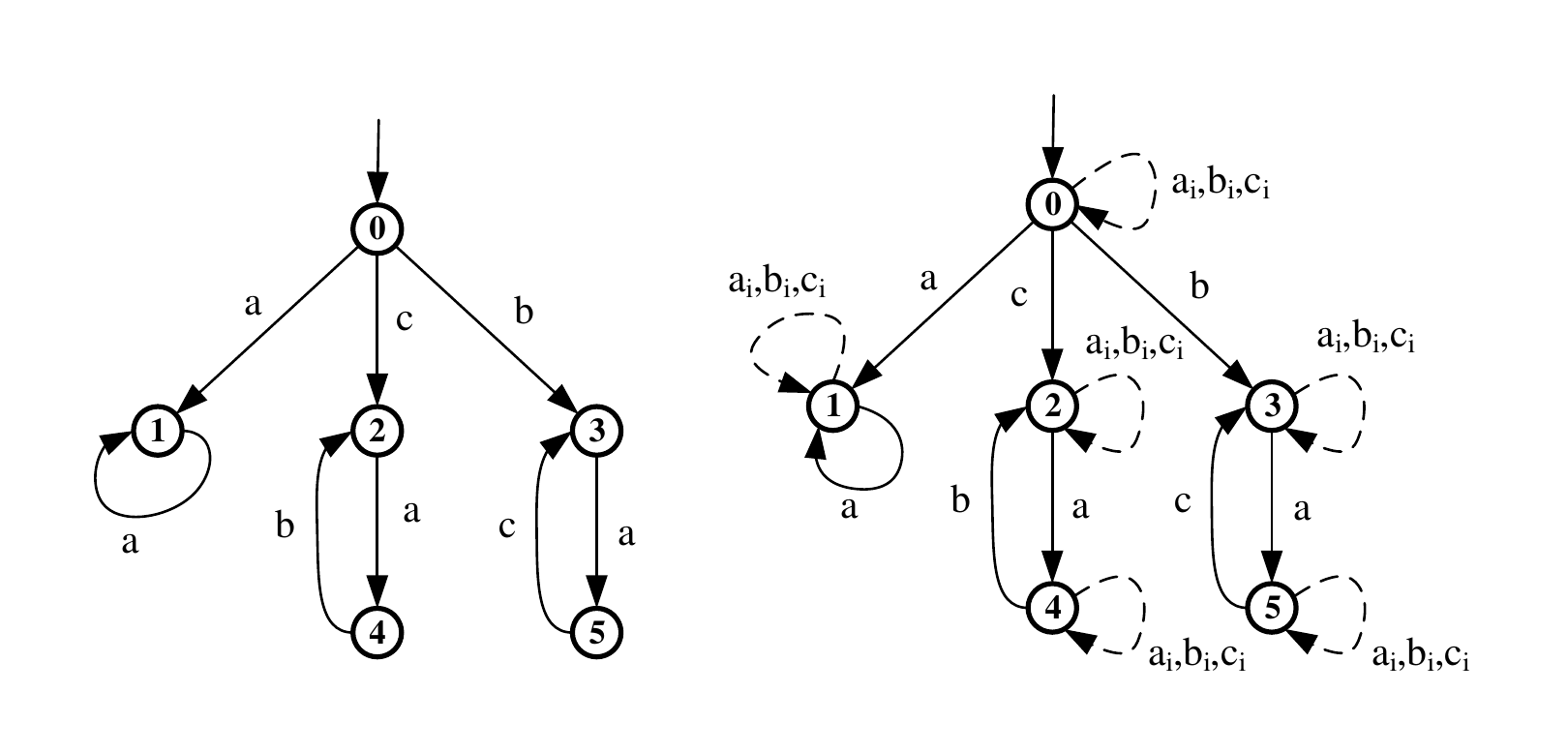}
\caption{DFA $G$ with $X_S=\{2, 3\}$ and its insertion automaton.}\label{DFA}
\end{figure}

\begin{myEg}
This example is used to illustrate the insertion sequence and the above descriptions.
Consider DFA $G$ illustrated on the left of Fig.~\ref{DFA}, where the set of states is $X=\{0, 1, 2, 3, 4 ,5\}$, the initial state is 0, and the set of observable events is $E_o=E=\{a, b, c\}$.
Considering the insertion mechanism, we have $E_{i}=\{a_{i}, b_{i}, c_{i}\}$.
For an observation $s=ba$, we have $|s|=2$, $s[1]=b$, and  $s[2]=a$.
Then, the insertion sequence for $s$ is $s_{I}(s)=s_{i}[1]s_{i}[2]$ where $s_{i}[1], s_{i}[2]\in E_{i}^{*}$, and correspondingly the modified observation after implementing $s_{I}(s)$ on $s$ is $s_{M}(s)=s_{i}[1]s[1]s_{i}[2]s[2]=s_{i}[1]bs_{i}[2]a$.
Suppose that $s_{i}[1]=c_{i}$ and $s_{i}[2]=a_{i}$. Then, the modified observation is $s_{M}(s)=c_{i}ba_{i}a$. Since the inserted event $e_{i}\in E_{i}^{*}$ is observationally equivalent to the actual event $e\in E$, the outside observer will observe sequence $cbaa$. Obviously, the insertion sequence $s_{i}[1]s_{i}[2]=c_{i}a_{i}$ is not proper since $cbaa\notin L(G)$.
Suppose that $s_{i}[1]=c_{i}a_{i}$ and $s_{i}[2]=a_{i}b_{i}$.
Then, the modified observation is $s_{M}(s)=c_{i}a_{i}ba_{i}b_{i}a$.
Since the inserted event $e_{i}\in E_{i}^{*}$ is observationally equivalent to the actual event $e\in E_{i}$, the outside observer will observe sequence $cababa$.
Obviously, the insertion sequence $s_{i}[1]s_{i}[2]=c_{i}a_{i}a_{i}b_{i}$ is proper since $cababa\in L(G)$.
\end{myEg}

\section{EI-ENFORCEABILITY AND EIC-ENFORCEABILITY}\label{enforceability}

This section starts with a motivating example for which the insertion mechanism in \cite{IF-opacity1,IF-opacity2,IF-opacity3} cannot enforce opacity, whereas the proposed extended insertion mechanism is able to do so.
The section discusses how to construct feasible extended insertion sequences (in order to avoid raising suspicion to the intruder),
and sustainable extended insertion sequences (in order to be able to continue inserting symbols without raising suspicion to the intruder when future activity is generated by the system).
To enforce opacity with a modified sequence that does not draw the intruder attention and can be maintained regardless of the continuation that is chosen by the system, this section introduces and analyzes the notion of desirable extended insertion sequences. Based on desirable extended insertion sequences, EI-enforceability is presented.
More specifically, if for every observation generated by a system, there exists a desirable extended insertion sequence, then the system is EI-enforceable.

In order to tackle opacity enforcement under event insertion constraints (EICs), the notions of EIC-feasible extended insertion sequence, EIC-sustainable extended insertion sequence, EIC-desirable extended insertion sequence, and EIC-enforceability are presented.
In the remainder of this section, we provide a motivating example and then state a number of definitions that are useful for verifying the two cases of extended insertion sequences that we mention above.

\subsection{Motivating example}

Reconsider the DFA $G$ illustrated on the left of Fig.~\ref{DFA}, where the set of secret states is $X_S=\{2, 3\}$. It is clear that the system is not current-state opaque since after event $b$ or $c$ occurs the state of the system is known to be $3\in X_S$ or $2\in X_S$.
If $c$ happens, the set of possible insertion event sequences is $\{\epsilon, b_ia_i(c_ia_i)^*\}$ according to method in \cite{IF-opacity1,IF-opacity2,IF-opacity3}. Consequently, the set of possible altered event \mbox{sequences} after the implementation of the insertion \mbox{sequences} is $\{c, b_ia_i(c_ia_i)^*c\}$.
Recall that from the perspective of an intruder, $e_i$ looks like $e$; thus the intruder observes the set of sequences $\{c, ba(ca)^*c\}=\{c, bac(ac)^*\}$. It is obvious that $\{c, bac(ac)^*\}\subseteq L_{cus}(G)$. Therefore, we cannot find an insertion sequence for $c$ to protect secret state 2 from being revealed. From the above descriptions, opacity of the DFA on the left of Fig.~\ref{DFA} cannot be enforced by the insertion mechanism proposed by Wu $et$ $al.$ \cite{IF-opacity1,IF-opacity2,IF-opacity3}.
On the other hand, we will see later on that an extended insertion mechanism can actually ensure opacity for this particular system.

\subsection{EI-enforceability and EIC-enforceability}
Assuming that the observation is
$s$ and $s[j]$ denotes the \mbox{$j$-th} observable event, where $j\in \{1, 2, ..., |s|\}$, we use $s_{EI}(s)=s_{bi}[1]s_{ai}[1]s_{bi}[2]s_{ai}[2]...s_{bi}[|s|]s_{ai}[|s|]$ to denote the corresponding extended insertion sequence of $s$, where $s_{bi}[j]$ and $s_{ai}[j]$ respectively denote the inserted strings before and after the $j$-th observable event.
Similar to insertion sequences, we use $e_i$ to denote an inserted event.

Given a DFA\footnote{Note that the more general case of an NFA under partial observation can also be treated by first constructing its observer (which would be a DFA under full observation) and then performing the analysis that we perform here using the observer as our DFA.}
$G=(X, E, f, x_0)$ with $E_i$ being the set of inserted events, we will also consider event insertion constraints (EICs): $E{_{i}^b}$ and $E{_{i}^a}$, where $E{_{i}^b}$ ($E{_{i}^a}$) represents the set of observable events that can be inserted before (after) an actual observed event.
We use $e_{bi}$ ($e_{ai}$) to denote that $e_{bi}\in E{_{i}^b}$ ($e_{ai}\in E{_{i}^a}$) can be inserted before (after) an actual observed event.
In the following parts, $E{_{i}^b}^*$ and $E{_{i}^a}^*$ are respectively used to represent the set of all finite-length strings of elements of $E{_{i}^b}$ and $E{_{i}^a}$ (the empty string $\epsilon$ is also a member of $E{_{i}^b}^*$ and $E{_{i}^a}^*$).

In order to analyze strings composed by actual observable events and inserted observable events, natural projections $P_i$ and $P_{ui}$, and mask $M_i$ are introduced.
Natural projection $P_i$ refers to actual events as unobservable, which is defined as $P_i(e_i)=e_i$ for $e_i\in E_i$, $P_i(e_{bi})=e_{bi}$ for $e_{bi}\in E{_{i}^b}$, $P_i(e_{ai})=e_{ai}$ for $e_{ai}\in E{_{i}^a}$,
and $P_i(e)=\epsilon$ for $e\in E$.
Natural projection $P_{ui}$ refers to inserted events as unobservable, which is defined as $P_{ui}(e)=e$ for $e\in E$, $P_{ui}(e_{bi})=\epsilon$ for $e_{bi}\in E{_{i}^b}$, $P_{ui}(e_{ai})=\epsilon$ for $e_{ai}\in E{_{i}^a}$,
and $P_{ui}(e_i)=\epsilon$ for $e_i\in E_i$.
Mask $M_i$ treats inserted events and actual events as indistinguishable, which is defined as $M_i(e_i)=e$ for $e_i\in E_i$, $M_i(e_{bi})=e$ for $e_{bi}\in E{_{i}^b}$, $M_i(e_{ai})=e$ for $e_{ai}\in E{_{i}^a}$,
and $M_i(e)=e$ for $e\in E$.
Given a string $s$, composed of inserted events and actual events, $P_{i}(s)$ erases all actual events in $s$, $P_{ui}(s)$ erases all inserted events in $s$, and $M_i(s)$ treats all events in $s$ (regardless of whether the event is actual or inserted) as actual events.

\begin{myDef}\label{feasible-insertion-sequence}(Feasible  extended insertion sequence)
Consider a DFA $G=(X, E, f, x_0)$ with the set of inserted events $E_i$. For an event sequence $s\in L(G)$, an extended insertion sequence $s_{EI}(s)=s_{bi}[1]s_{ai}[1]s_{bi}[2]s_{ai}[2]...$ $s_{bi}[|s|]s_{ai}[|s|]$ is said to be feasible for $s$ if
$$\forall n\in \{1, 2, ..., |s|\}\{(s'=s[1]s[2]...s[n]\in \overline{s})\}$$ we have $M_i(s_{EM}(s'))\in L(G)$, where $s_{EM}(s')=s_{bi}[1]s[1]s_{ai}[1]\\s_{bi}[2]s[2]s_{ai}[2]...s_{bi}[n]s[n]s_{ai}[n], s_{bi}[k], s_{ai}[k]\in E_i^{*}, k\in \{1, 2, ..., \\n\}$, and $n$ is a positive integer.
\end{myDef}

\begin{myDef}\label{sustainable-insertion-sequence}(Sustainable extended insertion sequence)
Given a DFA $G=(X, E, f, x_0)$ with $E_i$ and an event sequence $s\in L(G)$, an extended insertion sequence $s_{EI}(s)=s_{bi}[1]s_{ai}[1]$ $s_{bi}[2]s_{ai}[2]...s_{bi}[|s|]s_{ai}[|s|]$ is said to be sustainable for $s$ if:\\ (1) $s_{EI}(s)$ is feasible for $s$;\\
(2) $\forall t=t[1]t[2]...t[|t|]\in SX(s)$ with $SX(s)=\{t|st\in L(G)\}$, we have $\{\exists s_{EI}(t)=t_{bi}[1]t_{ai}[1]t_{bi}[2]t_{ai}[2]...t_{bi}[|t|]t_{ai}[|t|]\}$ s.t. $s_{EI}(st)=s_{EI}(s)s_{EI}(t)$ is feasible for $st$.
\end{myDef}

\begin{myDef}\label{desirable-insertion-sequence}(Desirable extended insertion sequence)
Given a DFA $G=(X, E, f, x_0)$ with $E_i$, set of secret states $X_S$, and an event sequence $s\in L(G)$, an extended insertion sequence $s_{EI}(s)=s_{bi}[1]s_{ai}[1]s_{bi}[2]s_{ai}[2]...s_{bi}[|s|]s_{ai}[|s|]$ is said to be desirable for $s$ if:\\ (1) $s_{EI}(s)$ is sustainable;\\
(2) $M_i(s_{EM}(s))\in L_{cs}(G)$.
\end{myDef}

\begin{myDef}\label{EI-enforceability}(EI-enforceability)
A DFA $G=(X, E, f, x_0)$, with $E_i$ and set of secret states $X_S$, is said to be EI-enforceable if $$\forall s\in L(G)\{(\exists s_{EI}(s)\}$$ such that $s_{EI}(s)$ is a desirable extended insertion sequence.
\end{myDef}

When considering insertion constraints on events, if $s_{bi}[k]$ and $s_{ai}[k]$ in Definition~\ref{feasible-insertion-sequence} satisfy $s_{bi}[k]\in E_{i}^{b*}$ and $s_{ai}[k]\in E_{i}^{a*}$, then the feasible extended insertion sequence $s_{EI}(s)$ would be EIC-feasible. It is not hard to see that the sustainable extended insertion sequence would be EIC-sustainable if the feasible extended insertion sequence in Definition~\ref{sustainable-insertion-sequence} is required to be EIC-feasible (with $s_{EI}(t)$ also satisfying the EICs); and the desirable extended insertion sequence would be EIC-desirable if the sustainable extended insertion sequence in Definition~\ref{desirable-insertion-sequence} is required to be EIC-sustainable.
Correspondingly, if the desirable extended insertion sequence in Definition~\ref{EI-enforceability} is required to be EIC-desirable, then a DFA would be EIC-enforceable.

\section{VERIFICATION FOR EI-ENFORCEABILITY AND EIC-ENFORCEABILITY}\label{enforceability-verification}

This section elaborates on how to built a verifier (EIC-verifier) that can be used to decide whether system opacity can be enforced via an extended insertion mechanism under no constraints (under event insertion constraints). Based on the obtained verifier (EIC-verifier), a necessary and sufficient condition to determine whether the system is EI-enforceable (EIC-enforceable) is presented.

\subsection{Verification for EI-enforceability}

\begin{myDef}\label{Insertion-automaton}(Insertion automaton)
Given a DFA $G=(X, E, f, x_0)$, the insertion automaton is given by $G^f=(X_f, E\cup E_{i}, f_{i}, x_{f, 0})$, where $X_f=X$, $x_{f, 0}=x_0$, and $f_i$ is defined as $f_i(x_f, e)=f(x_f, e)$ (denoted by solid lines) for $x_f\in X_f$ and $e\in E$, and $f_i(x_f, e_i)=x_f$ (denoted by dashed lines) for $x_f\in X_f$ and $e_i\in E_i$.
\end{myDef}

\begin{myDef}\label{Dashed-product}(Indicator automaton)
Given a DFA $G=(X, E, f, x_{0})$ and its insertion automaton $G^f=(X_f, E\cup E_{i},$ $f_{i}, x_{f, 0})$,
the indicator automaton is the dashed product of $G$ and $G^f$, denoted by
$G_{IA}=G\times _{d}G^f=(X_{IA}, E\cup E_{i}, f_{IA}, x_{IA,0})$, where\\
(1) $X_{IA}=\{(x, x_f)|x\in X, x_f\in X_f\}$ is the set of states;\\
(2) $E\cup E_{i}$ is the set of events, where $E_{i}$ is the set of inserted events and $e_i\in E_{i}$ is observationally equivalent to $e\in E$;\\
(3) $x_{IA,0}=(x_{0}, x_{f, 0})$ is the initial state;\\
(4) $f_{IA}$ is the transition function defined as follows:
(i) $f_{IA}((x, x_f), e)=(f(x, e), f_i(x_f, e))$ if both $f(x, e)$ and $f_i(x_f, e)$ are defined, denoted by solid lines, and (ii) $f_{IA}((x, x_f), e_i)=(f(x, e)$, $f_i(x_f, e_i))$ if both $f(x, e)$ and $f_i(x_f, e_i)$ are defined, denoted by dashed lines.
\end{myDef}

\begin{myLem}\label{enumerate-feasible-extended-insertion-sequence}
Consider a DFA $G=(X, E, f, x_{0})$. Let $G^f=(X_f, E\cup E_{i},$ $f_{i}, x_{f, 0})$ be the corresponding insertion automaton and $G_{IA}=(X_{IA}, E\cup E_{i}, f_{IA}, x_{IA,0})$ be the corresponding indicator automaton.
Indicator automaton $G_{IA}$ enumerates all feasible extended insertion sequences.\hfill $\square$
\end{myLem}

Due to space limitation, the proof here is omitted. For same reason, all proofs in the remainder of paper are also omitted. They will be made available in an extended version of the paper.

\begin{figure}[t]
\centering
\includegraphics[height=7.7cm]{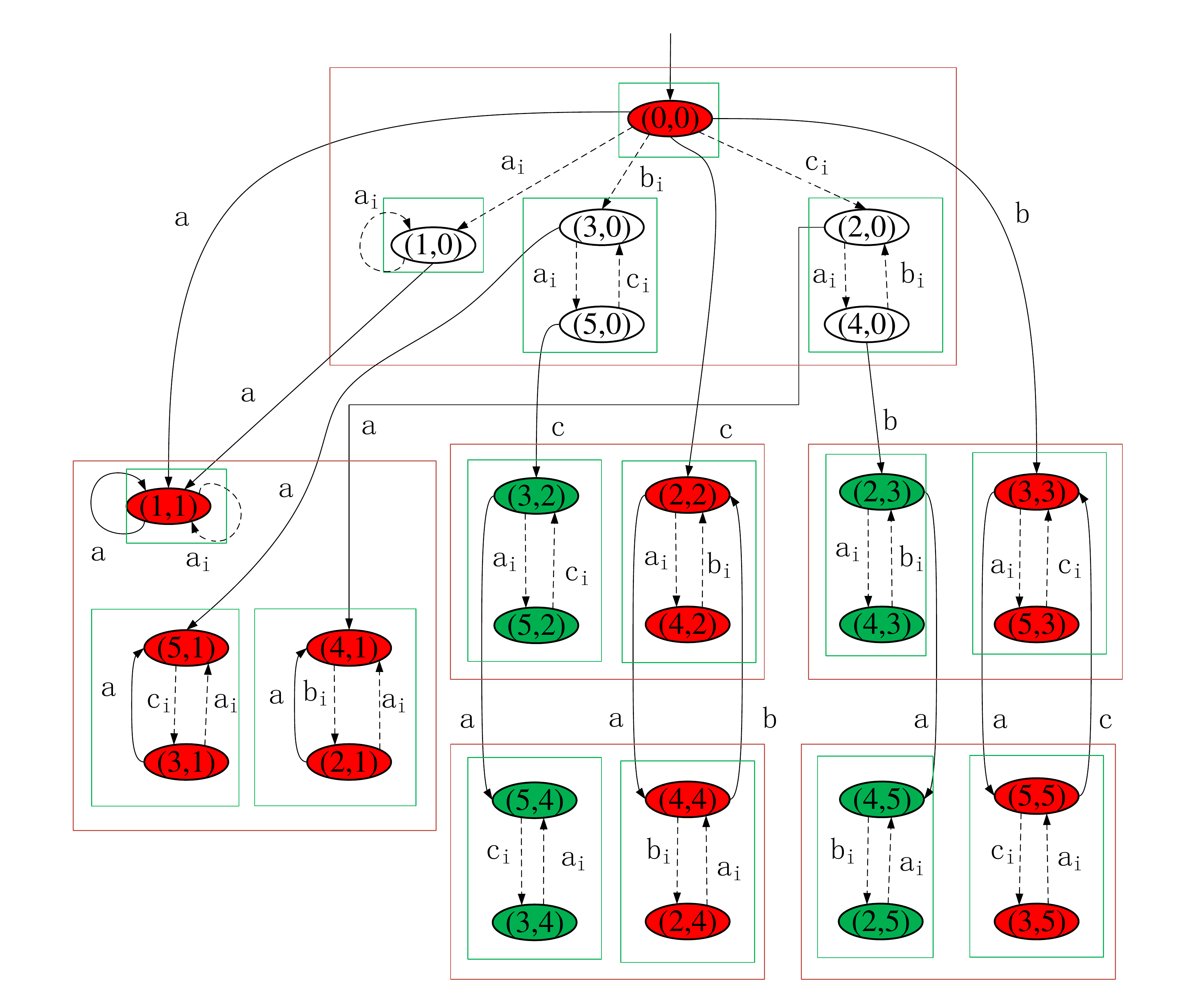}
\caption{Indicator automaton of the DFA on the left of Fig.~\ref{DFA}.}\label{indicator}
\end{figure}

\begin{myEg} From Definition~\ref{Insertion-automaton}, an insertion automaton is effectively obtained by inserting all observable events as self loops at each state of the given automaton.
As an example, consider the DFA on the left of Fig.~\ref{DFA}. After adding self-loops with all observable events at each state, the corresponding insertion automaton can be obtained, as illustrated on the right of Fig.~\ref{DFA}.
Following Definition~\ref{Dashed-product}, the indicator automaton is obtained as depicted in Fig.~\ref{indicator}.
\end{myEg}

Given an indicator automaton $G_{IA}=(X_{IA}, E\cup E_{i}, f_{IA}, x_{IA,0})=(X, E, f, x_{0})\times_{d}(X_f, E\cup E_{i}, f_{i}, x_{f, 0})$ with $X=\{x_1, x_2, ..., x_{|X|}\}$, the state space $X_{IA}$ can be partitioned into $|X|$ mutually disjoint subspaces: $X_{IA}=X_{IA}(x_{1})\cup X_{IA}(x_{2})\cup... \cup X_{IA}(x_{|X|})$, where $X_{IA}(x_{k})=\{x_{IA}=(x, x_f)\in X_{IA}|x_{f}=x_{k}\}$ is the set of first-level state subspaces and $k\in \{1, 2, ..., |X|\}$ is the index of state $x_k$.
Furthermore, the set of states (state subspace) $X_{IA}(x_{k})$
can be divided into $n_k$ mutually disjoint (note here that $n$ is a function of $k$) strongly connected components (SCCs): $X_{IA}(x_{k})=X_{IA}^{SCC_1}(x_k)\cup X_{IA}^{SCC_2}(x_k)\cup...\cup X_{IA}^{SCC_{n_{k}}}(x_k)$, where $X_{IA}^{SCC_{m_{k}}}(x_k)$ is a second-level subspace, $m_{k}\in\{1, 2, ..., n_k\}$ and $n_k$ is a positive integer denoting the number of SCCs obtained from $X_{IA}(x_{k})$.
In this paper, Tarjan Algorithm from \cite{identify-SCC} is used to obtain the SCCs of each subgraph of $G_{IA}$ with $X_{IA}(x_{k})$ being the set of nodes and \{$(x_{IA}, f_{IA}(x_{IA}, e_{i}))|x_{IA}\in X_{IA}(x_{k})\wedge f_{IA}(x_{IA}, e_{i})\in X_{IA}(x_{k})$ is defined\} being the set of directed edges.

\begin{myDef}\label{Trapping-SCCs}(Trapping SCC)
Let $X_{IA}^{SCC_{m_{k}}}(x_k)$ be a second-level state subspace of $G_{IA}=(X_{IA}, E\cup E_{i}, f_{IA},$ $x_{IA,0})=(X, E, f, x_{0})\times_{d}(X_f, E\cup E_{i},$ $f_{i}, x_{f, 0})$. The second-level subspace $X_{IA}^{SCC_{m_{k}}}(x_k)$ is said to be trapping if it satisfies the following two conditions:
(1) $\forall x_{IA}\in X_{IA}^{SCC_{m_{k}}}(x_k)$ and $\forall e\in T(x_{k})$, $f_{IA}(x_{IA}, e)$ is not defined; and (2) $\forall x_{IA}\in X_{IA}^{SCC_{m_{k}}}(x_k)$ and $\forall e_{i}\in E_{i}$, we have $f_{IA}(x_{IA}, e_{i})\in X_{IA}^{SCC_{m_{k}}}(x_k)$ or $f_{IA}(x_{IA}, e_{i})$ is not defined.
\end{myDef}

\begin{myDef}\label{Verifier}(Verifier)
Given an indicator automaton $G_{IA}=(X_{IA}, E\cup E_{i}, f_{IA}, x_{IA,0})=(X, E, f$, $x_{0})\times_{d}(X_f, E\cup E_{i},$ $f_{i}, x_{f, 0})$, a subgraph of $G_{IA}$, denoted by $G_{TV}=(X_{TV}, E\cup E_{i}, f_{TV}, x_{{TV},0})$, is said to be a verifier if trapping SCCs are iteratively pruned and only the part of the automaton that is accessible from the initial state is kept.
\end{myDef}

\begin{myEg} Consider the indicator automaton in Fig.~\ref{indicator}.
The whole state space in Fig.~\ref{indicator} can be partitioned into six first-level state subspaces since $G$ has six states. The first-level state subspaces are separately bounded in the figure by red lines.
A first-level state subspace can be categorized into several second-level subspaces that are bounded by green lines; this categorization depends on how many SCCs can be captured within the corresponding first-level state subspace.
Consider the second-level subspace $X_{IA}^{SCC}(4)=\{(3, 4), (5, 4)\}$ as an example.
It is obvious that (1) $\forall x_{IA}\in \{(3, 4), (5, 4)\}$ and $\forall e\in T(4)=\{b\}$, $f_{IA}(x_{IA}, e)$ is not defined (neither $f_{IA}((3, 4), b)$ is defined nor $f_{IA}((5, 4), b)$ is defined);
and (2) $\forall x_{IA}\in \{(3, 4), (5, 4)\}$ and $\forall e_{i}\in E_{i}$ we have $f_{IA}(x_{IA}, e_{i})\in \{(3, 4), (5, 4)\}$ or $f_{IA}(x_{IA}, e_{i})$ is undefined,
which means that the second-level subspace $X_{IA}^{SCC}(4)$ is trapping.
After iteratively pruning all trapping SCCs and taking accessible part from initial state, the verifier is obtained as shown in Fig.~\ref{indicator} if we exclude the SCCs filled with color green and their incoming events.
\end{myEg}

\begin{myDef}\label{Staying-nonblocking-SCC}(Staying nonblocking SCC in a verifier)
Consider an indicator automaton $G_{IA}=(X_{IA}, E\cup E_{i}, f_{IA}, x_{IA,0})=(X, E, f$, $x_{0})\times_{d}(X_f, E\cup E_{i},$ $f_{i}, x_{f, 0})$ and its verifier $G_{V}=(X_{V}, E\cup E_{i}, f_{V}, x_{{V},0})$.
The second-level subspace $X_{V}^{SCC_{m_{k}}}(x_k)$ is said to be staying nonblocking if $\forall e\in T(x_{k})$ \{$\exists x_{V}\in X_{V}^{SCC_{m_{k}}\uparrow}(x_k)$\{$f_{V}(x_{V}, e)$ is defined\}\}, where $X_{V}^{SCC_{m_{k}}\uparrow}(x_k)=\{X_{V}^{SCC_{m_{k}'}}(x_k)|\exists$ $x_{V}\in X_{V}^{SCC_{m_{k}}}(x_k)$, $x_{V}'\in X_{V}^{SCC_{m_{k}'}}(x_k)$, and $s_{bi}\in E_{i}^{*}$ \{$f_{V}(x_{V}, s_{bi})=x_{V}'$\}\}.
Let $X_{VSNB}$ denote the set of states that lie in staying nonblocking SCCs in $G_{V}$.
\end{myDef}

\begin{myLem}\label{enumerate-sustainable-extended-insertion-sequence}
Consider a DFA $G=(X, E, f, x_{0})$ and construct its indicator automaton $G_{IA}=(X_{IA}, E\cup E_{i}, f_{IA}, x_{IA,0})$ and verifier $G_{V}=(X_{V}, E\cup E_{i}, f_{V}, x_{{V},0})$.
Let $X_{VSNB}$ be the set of states that lie in staying nonblocking SCCs.
The verifier enumerates all sustainable extended insertion sequences as the modified sequences that can reach a state in $X_{VSNB}$.\hfill $\square$
\end{myLem}

\begin{myDef}\label{admissible-state}(Admissible state in a verifier)
Given a DFA $G=(X, E, f, x_0)$ with $X_S\subseteq X$ being the set of secret states, $G_{V}=(X_{V}, E\cup E_{i}, f_{{V}}, x_{{V}, 0})$ being its verifier, and $x_{VSNB}$ being its set of states that lie in staying nonblocking SCCs, a state $x_{VSNB}=(x, x_{f})\in X_{VSNB}$ is said to be admissible if $x\in (X\backslash X_S)$. The set of admissible states is denoted by $X_{VA}$.
\end{myDef}

For $x_{VSNB}=(x, x_f)\in X_{VSNB}$, $x$ is the dummy state that can be erroneously inferred by an intruder and $x_f$ is the genuine state of the system.
To protect the privacy of the system, the secret states cannot be exposed to the intruder.
Therefore, if the dummy state $x$ is not secret then the state $x_{VSNB}=(x, x_f)$ is admissible regardless of whether the genuine state $x_f$ is secret or not.

\begin{myTheo}\label{Theorem-EI-enforceable}
Let $G_{V}=(X_{V}, E\cup E_{i}, f_{V}, x_{{V},0})$ be the verifier of a DFA $G=(X, E, f, x_0)$ with $X_S\subseteq X$ being the set of secret states. The automaton $G$ is EI-enforceable if and only if $$\forall x_f\in X_f\{\exists x_{V}=(x, x_f)\in X_{VA}\},$$
where $X_{VA}$ is the set of admissible states of the verifier $G_V$.\hfill $\square$
\end{myTheo}

\begin{myEg} Consider the verifier $G_V$ in Fig.~\ref{indicator} (without the SCCs filled with color green and their incoming events).
Second-level subspace $X_{V}^{SCC}(0)=\{(0, 0)\}$ is staying nonblocking since for all $e\in T(0)=\{a, b, c\}$, we have that $f_{V}((0, 0), e)$ is defined ($f_{V}((0, 0), a)$, $f_{V}((0, 0), b)$, and $f_{V}((0, 0), c)$ are all defined).
For this example, all staying nonblocking SCCs are filled with color red as shown in Fig.~\ref{indicator}, and we have $X_{VSNB} = \{(0, 0),$ $(1, 1), (2, 1), (3, 1), (4, 1), (5, 1), (2, 2), (4, 2), (3, 3),(5, 3), (4, 4),$ $(2, 4), (5, 5), (3, 5)\}$.
The states $(2, 2)$ and $(2, 4)$ are not admissible while $(4, 1)$ and $(4, 2)$ are admissible. For states $(2, 2)$ and $(2, 4)$, dummy state 2 is secret implying that both of these states are not admissible. States $(4, 1)$ and $(4, 2)$ are admissible since dummy state 4 is not secret.
Overall, the set of admissible states is $X_{VA}=\{(0, 0), (1, 1),$ $(4, 1), (5, 1), (4, 2), (5, 3), (4, 4), (5, 5)\}$. Based on Theorem~\ref{Theorem-EI-enforceable}, the DFA on the left of Fig.~\ref{DFA} is EI-enforceable.
\end{myEg}

\subsection{Verification for EIC-enforceability}

This section concentrates on verification for EIC-enforceability, i.e., whether a nonopaque system with EICs can be turned into an opaque one via an extended insertion mechanism.
Similar to the verification for EI-enforceability, we construct an EIC-verifier, based on which a necessary and sufficient condition is presented to determine whether the considered system with EICs is EIC-enforceable.

\begin{myDef}\label{EIC-Insertion-automaton} (EIC-insertion automaton)
Given a DFA $G=(X, E, f, x_0)$ with EICs captured by $E{_{i}^b}$ and $E{_{i}^a}$, the EIC-insertion automaton is a DFA $G^{EIC}=(X_f, E\cup E{_{i}^b}\cup E{_{i}^a},$ $f_{i}, x_{f, 0})$, where
\begin{enumerate}
  \item $X_f=X\cup X_{a}\cup X_{b}\cup X_{ab}$ is the set of states, with
\begin{enumerate}
           \item $X_{a}=\{x_{a}|x\in X\}$
           \item $X_{b}=\{x_{b}|x\in X\}$
           \item $X_{ab}=\{x_{ab}|x\in X\}$.
         \end{enumerate}
  \item $x_{f, 0}=x_0$ is the initial state.
  \item the state transition function $f_i$ is defined as follows:\begin{enumerate}
           \item state transition function associated with actual observable events (denoted by solid lines):\begin{enumerate}
                    \item $f_i(x, e)=f(x, e)$ for $x\in X$ and $e\in E$.
                    \item $f_i(x_a, e)=f(x, e)$ for $x_{a}\in X_{a}$ and $e\in E$.
                    \item $f_i(x_b, e)=f(x, e)$ for $x_{b}\in X_{b}$ and $e\in E$.
                    \item $f_i(x_{ab}, e)=f(x, e)$ for $x_{ab}\in X_{ab}$ and $e\in E$.
                  \end{enumerate}
           \item state transition function associated with inserted observable events (denoted by dashed lines):\begin{enumerate}
                    \item $f_i(x, e_{ai})=x_{a}$ for $x\in X$ and $e_{ai}\in E{_{i}^a}$.
                    \item $f_i(x, e_{bi})=x_{b}$ for $x\in X$ and $e_{bi}\in E{_{i}^b}$.
                    \item $f_i(x_{a}, e_{ai})=x_{a}$ for $x_{a}\in X_{a}$ and $e_{ai}\in E{_{i}^a}$.
                    \item $f_i(x_{a}, e_{bi})=x_{ab}$ for $x_{a}\in X_{a}$ and $e_{bi}\in E{_{i}^b}$.
                    \item $f_i(x_{b}, e_{bi})=x_{b}$ for $x_{b}\in X_{b}$ and $e_{bi}\in E{_{i}^b}$.
                    \item $f_i(x_{ab}, e_{bi})=x_{ab}$ for $x_{ab}\in X_{ab}$ and $e_{bi}\in E{_{i}^b}$.
                  \end{enumerate}
         \end{enumerate}
\end{enumerate}
\end{myDef}

\begin{figure}[t]
  \centering
  \includegraphics[height=4.2cm]{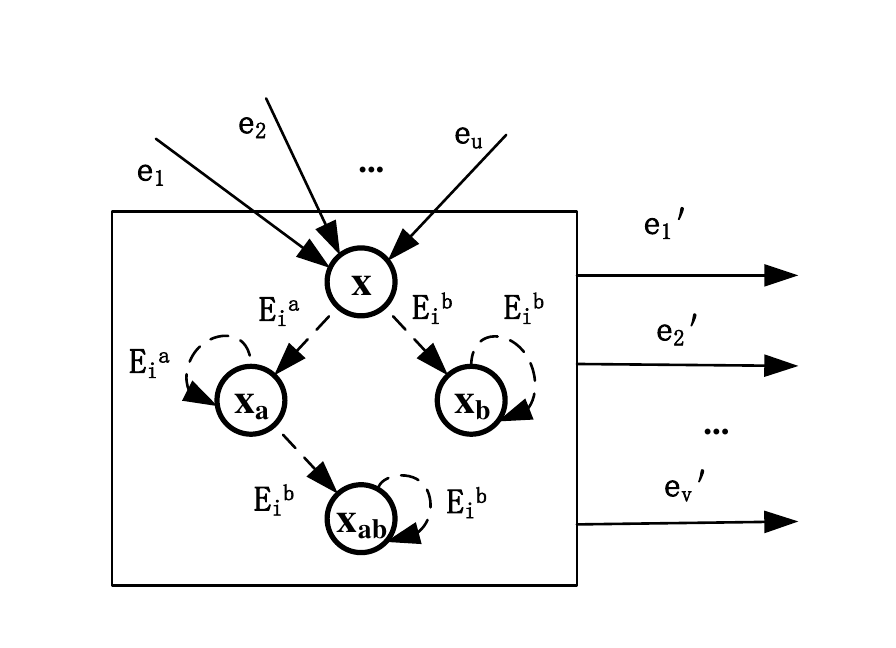}\\
  \caption{Illustration for EIC-insertion automaton.}\label{illustration-insertion-automaton-EIC}
\end{figure}

Suppose that the original system has a state $x$ with $F(x)=\{e_{1}, e_{2}, ..., e_{u}\}$ and $T(x)=\{e_{1}', e_{2}', ..., e_{v}'\}$, where $u$ and $v$ are two positive integers.
We explain \mbox{Definition~\ref{EIC-Insertion-automaton}} in Fig.~\ref{illustration-insertion-automaton-EIC}, where we simply use $E_{i}^{a}$ to denote that all $e_{ai}\in E_{i}^{a}$ can be inserted after and $E_{i}^{b}$ to denote that all $e_{bi}\in E_{i}^{b}$ can be inserted before.
In Fig.~\ref{illustration-insertion-automaton-EIC}, $x_{a}$ represents the system actual state $x$ reached by inserting virtual events (that can only be inserted after an actual observed event) after an observed event $e\in \{e_{1}, e_{2}, ..., e_{u}\}$;
$x_{b}$ represents the system actual state $x$ reached by inserting virtual events (that can only be inserted before an actual observed event) before an observed event $e'\in \{e_{1}', e_{2}', ..., e_{v}'\}$; and
$x_{ab}$ represents the system actual state $x$ reached by first inserting virtual events (that can only be inserted after an actual observed event) after an observed event $e\in \{e_{1}, e_{2}, ..., e_{u}\}$ and then inserting virtual events (that can only be inserted before an actual observed event) before an observed event $e'\in \{e_{1}', e_{2}', ..., e_{v}'\}$.
At all states $x$, $x_{a}$, $x_{b}$, and $x_{ab}$, event $e\in E$ satisfying $e\in T(x)$ can occur since the inserted event does not influence the system operation (refer to 3.(a) in Definition~\ref{EIC-Insertion-automaton});
at state $x$ event $e_{ai}\in E{_{i}^a}$ and $e_{bi}\in E{_{i}^b}$ can both be inserted (refer to 3.(b)(i) and (ii) in Definition~\ref{EIC-Insertion-automaton});
at state $x_{a}$ event $e_{ai}\in E{_{i}^a}$ and $e_{bi}\in E{_{i}^b}$ can both be inserted (refer to 3.(b)(iii) and (iv) in Definition~\ref{EIC-Insertion-automaton});
at state $x_{b}$ only event $e_{bi}\in E{_{i}^b}$ can be inserted (refer to 3.(b)(v) in Definition~\ref{EIC-Insertion-automaton});
and at state $x_{ab}$ only event $e_{bi}\in E{_{i}^b}$ can be inserted (refer to 3.(b)(vi) in Definition~\ref{EIC-Insertion-automaton}).

\begin{myRek}
Following the proposed extended insertion mechanism, an intruder can only observe state $x$ or $x_{a}$.
When the system is at initial state $x_{0}$, we only need to decide what to insert before the next observed event, which means that $x_{{0}_{a}}$ and $x_{{0}_{ab}}$ can be deleted in EIC-insertion automaton if the system cannot reach the initial state $x_{0}$ during subsequent operation (this holds if $F(x_{0})=\emptyset$), otherwise $x_{{0}_{a}}$ and $x_{{0}_{ab}}$ need to be retained (i.e., if the system can reach the initial state $x_{0}$ during subsequent operation which holds if \mbox{$F(x_{0})\neq \emptyset$)}.
\end{myRek}

\begin{figure}[t]
\centering
\includegraphics[height=7.8cm]{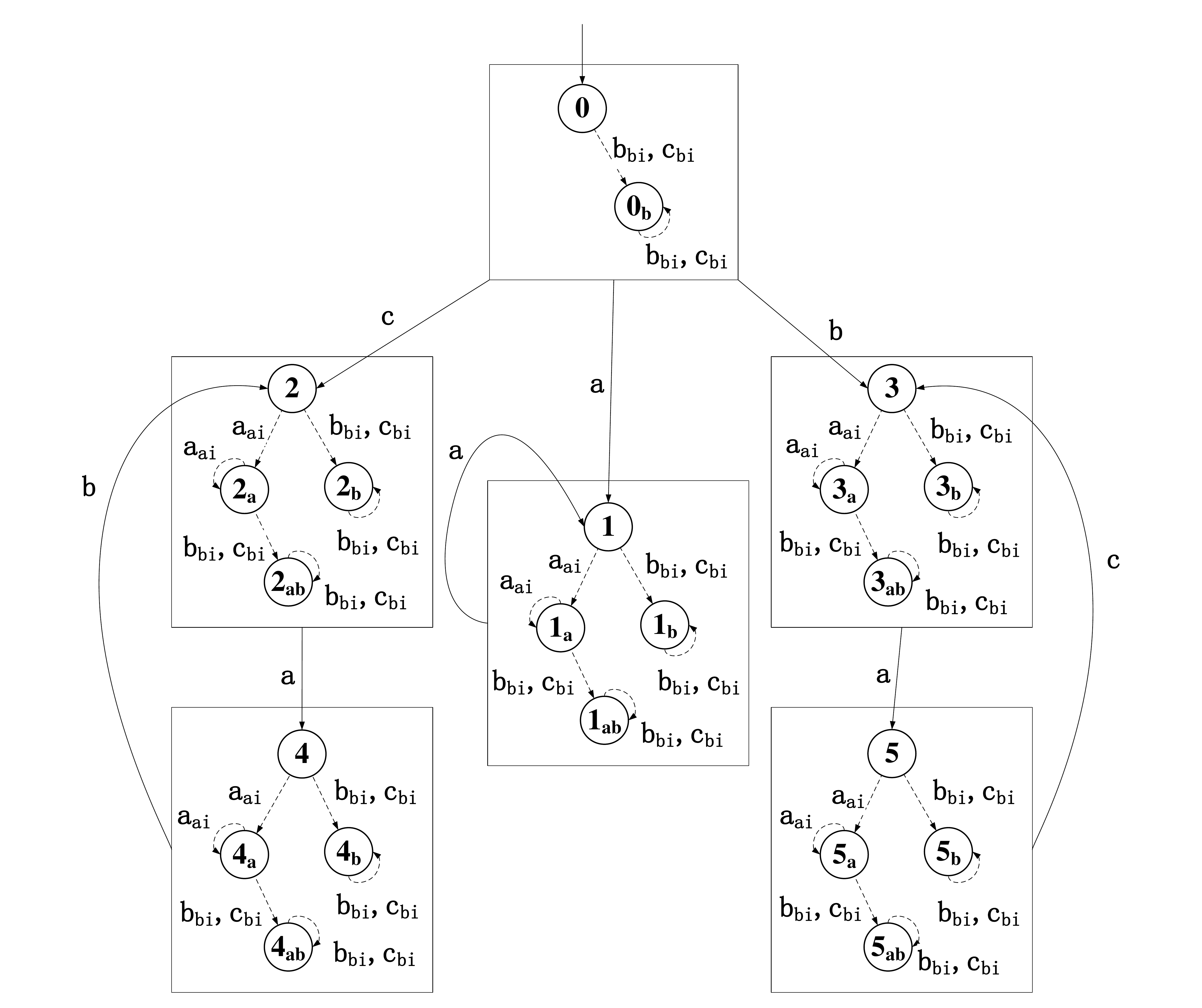}
\caption{EIC-insertion automaton of the DFA in Fig.~\ref{DFA} with $E_i^{b}=\{b_{bi}, c_{bi}\}$ and $E_i^{a}=\{a_{ai}\}$.}\label{EIC-insertion-automaton}
\end{figure}

\begin{myEg} Consider the DFA in Fig.~\ref{DFA}. Assuming that $b, c\in E$ are allowed to be inserted before an actual observed event, and $a\in E$  is allowed to be inserted after an actual observed event, then we have $E{_{i}^b}=\{b_{bi}, c_{bi}\}$ and $E{_{i}^a}=\{a_{ai}\}$.
Note that at initial state $x_0$ we have $F(x_0)=\emptyset$, implying that only $e_{bi}\in E{_{i}^b}$ can be inserted at $x_0$.
Based on Definition~\ref{EIC-Insertion-automaton}, the EIC-insertion automaton of the DFA in Fig.~\ref{DFA} with $E{_{i}^b}=\{b_{bi}, c_{bi}\}$ and $E{_{i}^a}=\{a_{ai}\}$ is obtained as illustrated in Fig.~\ref{EIC-insertion-automaton}.
Notice that an arc associated with an event $e\in E$ from a rectangle represents that event $e$ causes a transition from all states that belong in the rectangle to the state indicated. For instance, $f_{i}(0, a)=1$ and $f_{i}(0_{b}, a)=1$.
\end{myEg}

\begin{myDef}\label{EIC-Dashed-product}(EIC-indicator automaton)
Given a DFA $G=(X, E, f, x_{0})$ with $E{_{i}^b}$ and $E{_{i}^a}$ and its EIC-insertion automaton $G^{EIC}=(X_f, E\cup E{_{i}^b}\cup E{_{i}^a}, f_{i}, x_{f, 0})$,
the EIC-indicator automaton is the EIC-dashed product of $G$ and $G^{EIC}$, denoted by
$G_{EIA}=G\times _{EIC-d}G^{EIC}=(X_{EIA}, E\cup E{_{i}^b}\cup E{_{i}^a}, f_{EIA}, x_{EIA,0})$ where\\
(1) $X_{EIA}=\{(x, x_f)|x\in X, x_f\in X_f\}$ is the set of states;\\
(2) $E\cup E{_{i}^b}\cup E{_{i}^a}$ is the set of events, where $E{_{i}^b}$ and $E{_{i}^a}$ are respectively the set of events that can be inserted before and after an actual observed event, and $e_{bi}\in E{_{i}^b}$ and $e_{ai}\in E{_{i}^a}$ are observationally equivalent to $e\in E$;\\
(3) $x_{EIA,0}=(x_{0}, x_{f, 0})$ is the initial state;\\
(4) $f_{EIA}$ is the transition function defined as \mbox{follows}:
(i) $f_{EIA}((x, x_f), e)=(f(x, e), f_i(x_f, e))$ if both $f(x, e)$ and $f_i(x_f, e)$ are defined, denoted by solid lines, (ii) $f_{EIA}((x, x_f), e_{bi})=(f(x, e)$, $f_i(x_f, e_{bi}))$ if both $f(x, e)$ and $f_i(x_f, e_{bi})$ are defined, denoted by dashed lines, and (iii) $f_{EIA}((x, x_f), e_{ai})=(f(x, e), f_i(x_f, e_{ai}))$ if both $f(x, e)$ and $f_i(x_f, e_{ai})$ are defined, denoted by dashed lines.
\end{myDef}

\begin{myLem}\label{enumerate-EIC-feasible-extended-insertion-sequence}
Consider a DFA $G=(X, E, f, x_{0})$. Let $G^{EIC}=(X_f, E\cup E{_{i}^b}\cup E{_{i}^a},$ $f_{i}, x_{f, 0})$ be the corresponding EIC-insertion automaton and $G_{EIA}=(X_{EIA}, E\cup E{_{i}^b}\cup E{_{i}^a}, f_{EIA}, x_{EIA, 0})$ be the corresponding EIC-indicator automaton.
EIC-indicator automaton $G_{EIA}$ enumerates all EIC-feasible extended insertion sequences.\hfill $\square$
\end{myLem}

\begin{myDef}\label{EIC-Blocking-transient-SCCs}(EIC-trapping state)
Consider an EIC-indicator automaton $G_{EIA}=(X_{EIA}, E\cup E{_{i}^b}\cup E{_{i}^a}, f_{EIA}$, $x_{EIA,0})$.
A state $x_{EIA}\in X_{EIA}$ is EIC-trapping if it satisfies the following conditions:
(1) $\forall e\in E$, $f_{EIA}(x_{EIA}, e)$ is not defined;
(2) $\forall e_{bi}\in E{_{i}^b}$, $f_{EIA}(x_{EIA}, e_{bi})$ is not defined;
 and (3) $\forall e_{ai}\in E{_{i}^a}$, $f_{EIA}(x_{EIA}, e_{ai})$ is not defined.
\end{myDef}

\begin{myDef}\label{EIC-Verifier}(EIC-verifier)
Given an EIC-indicator automaton $G_{EIA}=(X_{EIA}, E\cup E{_{i}^b}\cup E{_{i}^a}, f_{EIA}, x_{EIA,0})$, a subgraph of $G_{EIA}$, denoted by $G_{EV}=(X_{EV}, E\cup E_{i}, f_{EV}, x_{{EV},0})$, is said to be an EIC-verifier if EIC-trapping states are iteratively pruned and only the part of the automaton that is accessible from the initial state is kept.
\end{myDef}

\begin{figure}[b]
\centering
\includegraphics[height=6cm]{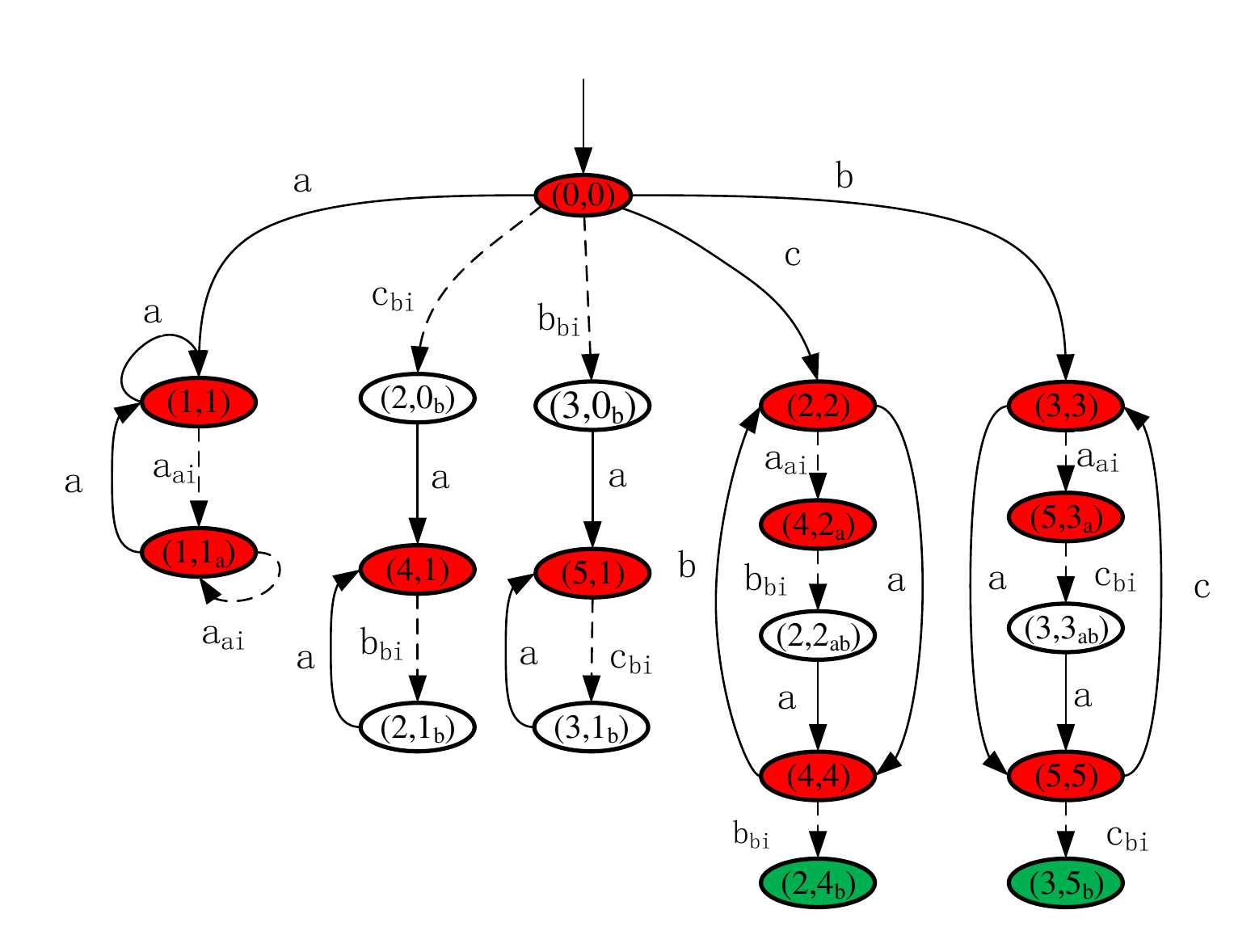}
\caption{EIC-indicator automaton of the DFA in Fig.~\ref{DFA} with $E{_{i}^b}=\{b_{bi}, c_{bi}\}$ and $E{_{i}^a}=\{a_{ai}\}$.}\label{EIC-indicator}
\end{figure}

\begin{myEg}
Consider the DFA in Fig.~\ref{DFA} with $E{_{i}^b}=\{b_{bi}, c_{bi}\}$ and $E{_{i}^a}=\{a_{ai}\}$. Following Definition~\ref{EIC-Dashed-product}, the corresponding EIC-indicator automaton is illustrated in Fig.~\ref{EIC-indicator}. According to Definition~\ref{EIC-Blocking-transient-SCCs}, states $(2, 4_{b})$ and $(3, 5_{b})$ are EIC-trapping.
Following Definition~\ref{EIC-Verifier}, after pruning them, the corresponding EIC-verifier can be obtained as illustrated in Fig.~\ref{EIC-indicator} if we exclude the states that are filled with the color green (and arrows from/to them).
\end{myEg}

We use $X_{EV}(x)=\{(x', x)\in X_{EV}\}\cup\{(x', x_{a})\in X_{EV}\}\cup \{(x', x_{b})\in X_{EV}\}\cup \{(x', x_{ab})\in X_{EV}\}$ to denote the set of states associated with the system actual state $x$ in an EIC-verifier.
Given an EIC-verifier $G_{EV}=(X_{EV}, E\cup E{_{i}^b}\cup E{_{i}^a}, f_{EV}, x_{EV,0})=(X, E, f, x_{0})\times_{EIC-d}(X_f, E\cup E{_{i}^b}\cup E{_{i}^a}, f_{i}, x_{f, 0})$ with $X=\{x_1, x_2, ..., x_{|X|}\}$, the state space $X_{EV}$ can be divided into $|X|$ mutually disjoint subspaces: $X_{EV}=X_{EV}(x_{1})\cup X_{EV}(x_{2})\cup... \cup X_{EV}(x_{|X|})$, where $X_{EV}(x_{k})$
is the set of first-level state subspaces and $k\in \{1, 2, ..., |X|\}$ is the index of \mbox{state} $x_k$.
Furthermore, the set of states (state subspace) $X_{EV}(x_{k})$ can be divided into 4 mutually disjoint
subspaces, i.e., $X_{EV}(x_{k})=X_{EV1}(x_k)\cup X_{EV2}(x_k)\cup X_{EV3}(x_k)\cup X_{EV4}(x_k)$, where $X_{EV1}(x_k)=\{x_{EV}=(x', x_{k})|x_{EV}\in X_{EV}(x_{k})\}$, $X_{EV2}(x_k)=\{x_{EV}=(x', x_{{k}_{a}})|x_{EV}\in X_{EV}(x_{k})\}$,
$X_{EV3}(x_k)=\{x_{EV}=(x', x_{{k}_{b}})|x_{EV}\in X_{EV}(x_{k})\}$, and
$X_{EV4}(x_k)=\{x_{EV}=(x', x_{{k}_{ab}})|x_{EV}\in X_{EV}(x_{k})\}$ are the sets of second-level state subspaces.

\begin{myDef}\label{EIC-nonBlocking-staying-SCCs}(Staying EIC-nonblocking state)
Consider an EIC-verifier $G_{EV}=(X_{EV}, E\cup E_{i}^{b}\cup E_{i}^{a}, f_{EV}, x_{{EV},0})$.\\
(1) A state $x_{EV}\in X_{EV1}(x_{k})$ is type-1 staying EIC-nonblocking if $\forall e\in T(x_{k})\{\exists x_{EV}' \in \{x_{EV}\}\cup x_{EV}^{\rightarrow}\{f_{EV}(x_{EV}', e)$ is defined\}\}, where $x_{EV}^{\rightarrow}=\{x_{EV}''\in X_{EV3}(x_{k})|\exists s_{bi}\in E_{i}^{b*}\{f_{EV}(x_{EV}, s_{bi})=x_{EV}''\}\}$ denotes the set of follow-on states of $x_{EV}$.\\
(2) A state $x_{EV}\in X_{EV2}(x_{k})$ is type-2 staying EIC-nonblocking if $\forall e\in T(x_{k})\{\exists x_{EV}' \in \{x_{EV}\}\cup x_{EV}^{\uparrow} \{f_{EV}(x_{EV}', e)$ is defined$\}\}$, where $x_{EV}^{\uparrow}=\{x_{EV}''\in X_{EV4}(x_{k})|\exists s_{bi}\in E_{i}^{b*}\{f_{EV}(x_{EV}, s_{bi})=x_{EV}''\}\}$ denotes the set of follow-on states of $x_{EV}$.\\
(3) Let $X_{EVNB}$ denote the set of type-1 and type-2 staying EIC-nonblocking states in $G_{EV}$.
\end{myDef}

\begin{myLem}\label{enumerate-EIC-sustainable-extended-insertion-sequence}
Consider a DFA $G=(X, E, f, x_{0})$. Let $G_{EV}=(X_{EV}, E\cup E{_{i}^b}\cup E{_{i}^a}, f_{EV}, x_{{EV},0})$ be the EIC-verifier and $X_{EVNB}$ be the set of type-1 and type-2 staying EIC-nonblocking states in $G_{EV}$. The EIC-verifier enumerates all EIC-sustainable extended insertion sequences as the sequences that reach a state in $X_{EVNB}$.\hfill $\square$
\end{myLem}

\begin{myDef}\label{EIC-admissible-state}(EIC-admissible state in an EIC-verifier)
Consider a DFA $G=(X, E, f, x_0)$ with its set of secret states $X_S\subseteq X$, event insertion constraints $E{_{i}^b}$ and $E{_{i}^a}$, and EIC-verifier $G_{EV}=(X_{EV}, E\cup E{_{i}^b}\cup E{_{i}^a}, f_{EV}, x_{{EV},0})$. A state $x_{EV}=(x, x')\in X_{EVNB}$  is said to be EIC-admissible if $x\in (X\backslash X_S)$, where $X_{EVNB}$ is the set of type-1 and type-2 staying EIC-nonblocking states. Let $X_{EVA}$ denote the set of EIC-admissible states in $G_{EV}$.
\end{myDef}

\begin{myTheo}\label{Theorem-EIC-enforceable}
Let $G_{EV}=(X_{EV}, E\cup E{_{i}^b}\cup E{_{i}^a}, f_{EV}, x_{{EV},0})$ be the EIC-verifier of a DFA $G=(X, E, f, x_0)$ with its set of secret states $X_S\subseteq X$ and event insertion constraints $E{_{i}^b}$ and $E{_{i}^a}$. The automaton $G$ is EIC-enforceable if and only if $$\forall x\in X \{\exists x_{EV}\in X_{EVA}\},$$
where $X_{EVA}$ is the set of EIC-admissible states of $G_{EV}$.\hfill $\square$
\end{myTheo}

\begin{myEg}
As an example, consider the EIC-verifier illustrated in Fig.~\ref{EIC-indicator} (without the states filled with color green and their incoming events). We have six first-level subspaces as follows: $X_{EV}(0)=\{(0, 0), (2, 0_{b}), (3, 0_{b})\}, X_{EV}(1)=\{(1, 1), (4, 1), (5, 1), (1, 1_{a}), (2, 1_{b}), (3, 1_{b})\}, X_{EV}(2)=\{(2, 2), (4, 2_{a}), (2, 2_{ab})\}, X_{EV}(3)=\{(3, 3), (5, 3_{a}), (3, 3_{ab})\}, \\X_{EV}(4)=\{(4, 4)\}$, and $X_{EV}(5)=\{(5, 5)\}$.
According to Definition~\ref{EIC-nonBlocking-staying-SCCs},
for state $x_{EV}=(4, 1)\in X_{EV1}(1)$, we have $x_{EV}^{\rightarrow}=\{(2, 1_{b})\}$, which indicates that $(4, 1)$ is type-1 staying $(1, 1)$-nonblocking since for all $e\in T(1)=\{a\}$, there exists $x_{EV}'=(2, 1_{b})\in x_{EV}^{\rightarrow}\cup \{x_{EV}\}=\{(4, 1), (2, 1_{b})\}$ such that
$f_{EV}(x_{EV}', e)$ is  defined.
According to Definition~\ref{EIC-nonBlocking-staying-SCCs},
for state $x_{EV}=(4, 2_{a})\in X_{EV2}(2)$, we have $x_{EV}^{\uparrow}=\{(2, 2_{ab})\}$, which indicates that $(4, 2_{a})$ is type-2 staying EIC-nonblocking since for all $e\in T(2)=\{a\}$, there exists $x_{EV}'=(2, 2_{ab})\in x_{EV}^{\uparrow}\cup \{x_{EV}\}=\{(4, 2_{a}), (2, 2_{ab})\}$ such that $f_{EV}(x_{EV}', e)$ is  defined.
In conclusion, states $(0, 0), (1, 1), (4, 1), (5, 1), (2, 2), (3, 3), (4, 4)$ and $(5, 5)$ are type-1 staying EIC-nonblocking; and $(1, 1_{a}), (4, 2_{a})$ and $(5, 3_{a})$ are type-2 staying EIC-nonblocking (these states are filled with color red as shown in Fig.~\ref{EIC-indicator}).
We use $X_{EVNB}$ to denote the set of type-1 and type-2 staying EIC-nonblocking states in an EIC-verifier; in this example, we have
$X_{EVNB} = \{(0, 0), (1, 1), (4, 1), (5, 1), (2, 2), (3, 3), (4, 4), (5, 5), (1, 1_{a}), \\(4, 2_{a}), (5, 3_{a})\}.$ There are nine EIC-admissible states in total, i.e., $X_{EVA}=\{(0, 0), (1, 1), (4, 1), (5, 1), (1, 1_{a}), (4, 2_{a}),\\ (5, 3_{a}), (4, 4), (5, 5)\}$ according to Definition~\ref{EIC-admissible-state}.
According to Theorem~\ref{Theorem-EIC-enforceable}, the system is EIC-enforceable.
\end{myEg}


\section{CONCLUSIONS AND FUTURE WORKS}\label{conclusions}

In this paper, we have introduced an extended insertion mechanism in order to enforce current-state opacity in discrete event systems that are modeled as fully observable DFAs. This is not a restriction as the approach can be extended to partially observable NFAs by first constructing their observer (which is a fully observable DFA).
Event insertion constraints (EICs) have been also considered in order to address practical limitations.
EI-enforceability of a system is presented to decide whether the opacity of the system can be enforced by the proposed mechanism, and EIC-enforceability of a system with EICs is presented to decide whether opacity of the system can be enforced by the proposed mechanism.
To verify EI-enforceability or EIC-enforceability, an appropriate verifier or EIC-verifier is constructed, and necessary and sufficient conditions are developed.
In the future, we will investigate how to enforce opacity of a system with other constraints via the proposed extended insertion mechanism.

%
%
%


\begin{thebibliography}{99}
\bibitem{opacity-presented}
L. Mazar\'{e}, ``Using unification for opacity properties,'' Proceedings of the 4th Working Group of International Federation for Information Processing, vol. 7, pp. 165--176, Oct. 2004.

\bibitem{Introduction-DES}
C. Cassandras, S. Lafortune, Introduction to Discrete Event Systems, 2nd ed. New York, NY, USA: Springer-Verlag, 2008.

\bibitem{chris-book}
C.~N. Hadjicostis, Estimation and Inference in Discrete Event Systems, Springer International Publishing, 2020.
\bibitem{tong-2017-v}
Y. Tong, Z. W. Li, C. Seatzu, A. Giua, ``Verification of state-based opacity using Petri nets,'' IEEE Transactions on Automatic Control, vol. 62, no. 6, pp. 2823--2837, Jun. 2017.

\bibitem{saboori2011verification1}
A.~Saboori, C.~N. Hadjicostis, ``Verification of infinite-step opacity and
  complexity considerations,'' IEEE Transactions on Automatic Control,
  vol.~57, no.~5, pp.~1265--1269, May 2012.

\bibitem{two-opacity-verify}
X. Yin, S. Lafortune, ``A new approach for the verification of infinite-step and $K$-step opacity using two-way observers,'' Automatica, vol. 80, pp. 162--171, Jun. 2017.

\bibitem{chris-2014}
A. Saboori, C. N. Hadjicostis, ``Current-state opacity formulations in probabilistic finite automata,'' IEEE Transactions on Automatic Control, vol. 59, no. 1, pp. 120--133, Jan. 2014.

\bibitem{chris-planning}
C. N. Hadjicostis, ``Trajectory planning under current-state opacity constraints,'' Proceedings of International Federation of Automatic control, vol. 51, no. 7, pp. 337--342, 2018.

\bibitem{Falcone}
Y. Falcone, H. Marchand, ``Enforcement and validation (at runtime) of various notions of opacity,'' Discrete Event Dynamic Systems, vol. 25, no. 4, pp. 531--570, Dec. 2015.

\bibitem{tong-2018-e}
Y. Tong, Z. W. Li, C. Seatzu, A. Giua, ``Current-state opacity enforcement in Discrete Event Systems under incomparable observations,'' Discrete Event Dynamic Systems, vol. 28, no. 2, pp. 161--182, Jun. 2018.

\bibitem{SC-opacity1}
J. Dubreil, P. Darondeau, H. Marchand, ``Supervisory control for opacity,'' IEEE Transactions on Automatic Control, vol. 55, no. 5, pp. 1089--1100, May 2010.

\bibitem{SC-opacity2}
S. Takai, Y. Oka, ``A formula for the supremal controllable and opaque
sublanguage arising in supervisory control,'' Journal of Control, Measurement, and System Integration, vol. 1, no. 4, pp. 307--312, 2008.

\bibitem{SC-opacity3}
A. Saboori, C. N. Hadjicostis, ``Opacity-enforcing supervisory strategies via
state estimator constructions,'' IEEE Transactions on Automatic Control, vol. 57, no. 5, pp. 1155--1165, May 2012.

\bibitem{dynamic-observability}
F. Cassez, J. Dubreil, H. Marchand, ``Synthesis of opaque systems with static and dynamic masks,'' Formal Methods in System Design, vol.~40, no.~1, pp. 88--115, Jan. 2012.

\bibitem{IF-opacity1}
Y.-C. Wu, S. Lafortune, ``Enforcement of opacity properties using insertion functions,'' Proceedings of 51st IEEE Conference on Decision and Control, Maui, Hawaii, USA, pp. 6722--6728, 2012.

\bibitem{IF-opacity2}
Y.-C. Wu, S. Lafortune, ``Synthesis of insertion functions for enforcement of opacity security properties,'' Automatica, vol. 50, no. 5, pp. 1336--1348, May 2014.

\bibitem{IF-opacity3}
Y.-C. Wu, S. Lafortune, ``Synthesis of optimal insertion functions for opacity enforcement,'' IEEE Transactions on Automatic Control, vol. 61, no. 3, pp. 571--584, Mar. 2016.

\bibitem{EIF-opacity}
C. Keroglou, S. Lafortune, ``Verification and synthesis of embedded insertion functions for opacity enforcement,'' Proceedings of 56th IEEE Conference on Decision and Control, pp. 4217--4223, 2017.

\bibitem{MIF-opacity}
C. Keroglou, L. Ricker, S. Lafortune, ``Insertion functions with memory for opacity enforcement,'' Proceedings of International Federation of Automatic control, vol. 51, no. 7, pp. 394--399, 2018.

\bibitem{PPIF-opacity}
Y. Ji, Y.-C. Wu, S. Lafortune, ``Enforcement of opacity by public and private insertion functions,'' Automatica, vol. 93, no. 7, pp. 369--378, Jul. 2018.

\bibitem{EF-1}
Y. Ji, X. Yin, S. Lafortune, ``Opacity enforcement using nodeterministic publicly-known edit functions,''
IEEE Transactions on Automatic Control, vol. 64, no. 10, pp. 4369--4376, Oct. 2019.


\bibitem{identify-SCC}
R. Tarjan, ``Depth-first search and linear graph algorithms,'' SIAM Journal on Computing, vol. 1, no. 2, pp. 146-160, Jun. 1972.
\end{thebibliography}
\end{document}